\documentclass[letter,twocolumn]{jpsj3}
\usepackage{txfonts}
\usepackage[dvipdfmx]{graphicx,color}
\usepackage{ulem}

\begin{document}
\title{Field-selective classical spin liquid and magnetization plateaus on kagome lattice}
\author{Kunio Tokushuku$^1$\thanks{tokushuku@hosi.phys.s.u-tokyo.ac.jp},
Tomonari Mizoguchi$^2$,
and Masafumi Udagawa$^{3,4}$}
\inst{$^1$Department of Physics, University of Tokyo, Tokyo, 113-0033, Japan \\
$^2$Department of Physics, University of Tsukuba, Tsukuba, Ibaraki 305-8571, Japan \\
$^3$Department of Physics, Gakushuin University, Mejiro, Toshima-ku, Tokyo 171-8588, Japan \\
$^4$Max-Planck-Institut f\"{u}r Physik komplexer Systeme, 01187 Dresden, Germany \\
}

\abst{We obtain a classical spin liquid (CSL) phase by applying a magnetic field to the $J_1$-$J_2$-$J_3$ Ising model on a kagome lattice. As we proved in the previous study [Phys. Rev. Lett. {\bf 119}, 077207 (2017)], this model realizes one species of CSL, the hexamer CSL, at the zero magnetic field, which consists of macroscopically degenerate spin configurations with mixed total magnetization, $M$. The magnetic field selects its subset, which can be mapped to a trimer covering of the dual lattice and forms a magnetization plateau of $M=1/9$.
In addition to this CSL, we find two other magnetization plateaus at $M=5/9$ and $17/27$, which are ascribed to the ``multimer" superstructures on a dual lattice.}

\maketitle
\textit{Introduction.-}
Realization of quantum spin liquid (QSL) is a central problem of condensed matter physics\cite{ANDERSON1973153,balents2010balents,RevModPhys.89.025003}.
Geometrical frustration is considered as one essential ingredient to realize this phase, and intensive efforts have been focused on the search of QSL in materials composed of triangular or tetrahedral basis units.
Theoretically, although the existence of QSL phases is established in a number of solvable models\cite{PhysRevLett.61.2376,KITAEV20032,kitaev2006anyons}, it still remains a difficult task to identify QSL ground state in a specific model, such as
the antiferromagnetic Heisenberg model on a kagome lattice~\cite{PhysRevB.45.12377,PhysRevB.46.14201,PhysRevLett.81.2356,PhysRevB.63.014413,PhysRevLett.89.137202,PhysRevB.68.214415,PhysRevLett.97.207204,PhysRevB.76.180407,Yan1173,PhysRevB.83.100404,doi:10.1143/JPSJ.80.053704,PhysRevLett.115.267209,PhysRevX.7.031020}.

One promising strategy to find QSL may be to focus on its high-temperature precursor, classical spin liquid (CSL).
At high temperatures, the QSL phase is sometimes preceded by cooperative paramagnetic states, composed of a degenerate assembly of classical moments under strong local constraints.
When the temperature decreases, CSL is gradually turned into QSL, as the quantum coherency develops.
The nature of QSL crucially depends on the basic characters of high-temperature CSL, such as the type of geometrical unit, the rule of local constraint, and so on.
This viewpoint, in turn, implies the possibility of engineering QSL with desirable properties by controlling its precedent CSL.

For this purpose, the application of a magnetic field provides a simple but practical method to control the local constraint of CSL.
It is particularly promising if a CSL state consists of degenerate configurations with different total magnetizations.
In this case, a magnetic field selects a part of the degenerate assembly and gives rise to a new CSL with different local constraints.
The transformation from spin ice to kagome ice~\cite{doi:10.1143/JPSJ.71.2365,isakov2004magnetization} gives an example of this mechanism, and it is experimentally well confirmed~\cite{matsuhira2002new,PhysRevLett.90.207205}.
The kagome ice state forms a $1/3$-magnetization plateau of Dy$_2$Ti$_2$O$_7$,
and it exhibits fertile phenomenology in thermodynamic and dynamical properties, which are absent in the original spin ice state\cite{doi:10.7566/JPSJ.82.073707,PhysRevB.90.144428,Lhotel:2018aa,PhysRevB.80.140409,PhysRevLett.106.207202,PhysRevLett.97.257205,PhysRevLett.90.207205,Carrasquilla:2015aa,PhysRevB.93.144402,PhysRevLett.119.227204,PhysRevLett.111.036602,doi:10.1142/S2010324715400044}.

\begin{figure*}[t]
\begin{center}
\includegraphics[clip, width=\linewidth]{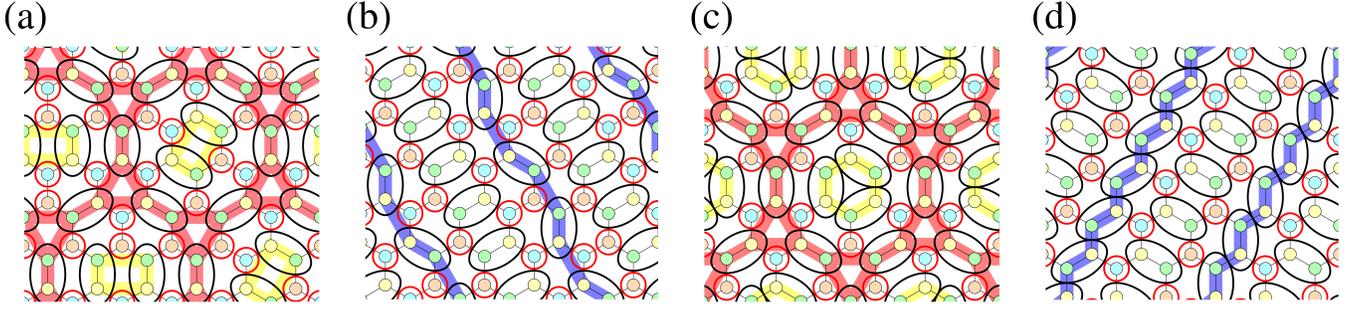}
\end{center}
\caption{(Color online) Typical charge configurations of
(a) the $17/27$ plateau with a kagome network,
(b) the $17/27$ plateau with a domain-wall structure,
(c) the $5/9$ plateau with a kagome network,
and (d) the $5/9$ plateau with a domain-wall structure.}
\label{fig:configuration}
\end{figure*}

In this paper, we focus on one class of CSL, which we named a hexamer CSL~\cite{PhysRevLett.119.077207}, realized in the $J_1$-$J_2$-$J_3$ Ising model on the kagome lattice.
This hexamer CSL is composed of the clusters of same-sign gauge charges, involving the configurations with mixed values of magnetization at the zero magnetic field.
For this state, we examine the state selection by the magnetic field, and obtained a magnetization plateau at $M=1/9$, where a new CSL state is stabilized.
This CSL state consists of a submanifold of the hexamer CSL, and it can be described as a trimer covering.
This picture turns out to give a simple starting point to understand the structure of the original hexamer CSL.
In addition, in the higher magnetic field, we find two more nontrivial magnetization plateaus at $M=17/27$ and $5/9$, attributed to superstructures of ``multimers", which are schematically shown in Fig.~\ref{fig:configuration}.

Below, we first introduce the language of dimers and monomers,
and clarify the origin of plateaus at $M=17/27$ and $5/9$ as a simple application of this language.
After that, we will address the $M=1/9$ plateau, and discuss its relation to the zero-field hexamer CSL.

\begin{figure}[b]
\begin{center}
\includegraphics[clip, width=\hsize]{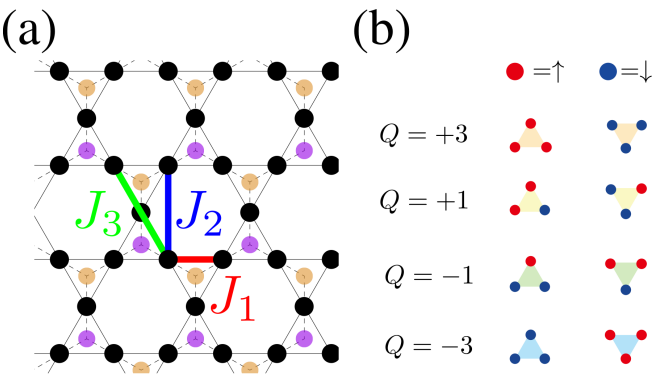}
\caption{(Color online) (a) Kagome $J_1$-$J_2$-$J_3$ Ising model.
The black circles denote the sites where Ising spins are located.
The sites on a dual honeycomb lattice with the sign factor $\eta_p = +1$ ($-1$)
are represented by the purple (orange) dots.
(b) The definition of charge variables.
The colors correspond to the values of charges $Q_p=+3$ (orange), $+1$ (yellow),
$-1$ (green), $-3$ (blue).}
\label{fig:model}
\end{center}
\end{figure}
{\it Model.-}
We consider the $J_1$-$J_2$-$J_3$ Ising model on the kagome lattice in a magnetic field:
\begin{align}
\mathcal{H}=& J_1 \sum_{\langle i,j \rangle_{\rm{n.n.}}} \sigma^{z}_i\sigma^{z}_j
+J_2 \sum_{ \langle i,j \rangle_{\rm{2nd}}} \sigma^{z}_i\sigma^{z}_j +J_3\sum_{\langle i,j \rangle_{\rm{3rd}}} \sigma^{z}_i\sigma^{z}_j - h\sum_{i}\sigma^{z}_i.
\label{eq:hamiltonian_kagome}
\end{align}
Here, $i$ and $j$ denote sites on the kagome lattice,
$\sigma^{z}_i = \pm1$ is the Ising spin variable on the site $i$,
and $\langle, \rangle_{\rm{n.n.}}$, $\langle , \rangle_{\rm{2nd}}$, and $\langle , \rangle_{\rm{3rd}}$
denote, respectively, the nearest-neighbor, the second-neighbor, and the third-neighbor pairs of sites;
see Fig.~\ref{fig:model}(a) for their definitions.

In this paper, we focus on the case of $J_1=1$ and $J_2= J_3=J$, with small positive $J$: $0 < J \lesssim 0.2$. In this case, the Hamiltonian can be written in a form of interacting charges~\cite{PhysRevB.88.100402,rau2016spin,PhysRevB.94.104416,PhysRevLett.119.077207,mizoguchi2018magnetic,PhysRevB.100.134415}.
The charge variable, $Q_p$, can be defined at each triangle
$p$ such that
\begin{align}
Q_p=\eta_p S_p, \label{eq:defcharge}
\end{align}
where
$S_p= \sum_{i\in p} \sigma^{z}_i$ is the total spin on a triangle $p$, and
$\eta_p=+1 (-1)$ for $p \in \bigtriangleup$ ($\bigtriangledown$),
representing the orientation of triangles.

The charges live on a dual lattice of the kagome lattice, namely, a honeycomb lattice [Fig.~\ref{fig:model}(a)].
They take the values, $Q_p=+3,+1,-1,-3$ [Fig.~\ref{fig:model}(b)].
Using $Q_p$, we can rewrite the Hamiltonian as
\begin{align}
\mathcal{H}= & \left( \frac{1}{2} -J \right) \sum_{p}Q_p^2-J \sum_{\langle p,q \rangle}Q_pQ_q
- \frac{h}{2}\sum_p\eta_pQ_p +C, \notag \\
\label{eq:hamiltonian_kagome2}
\end{align}
where $C=\frac{3}{2}(J-1)N_p$ is a constant term, with $N_p$ being the number of triangles.
The first, the second, and the third terms of Eq.~(\ref{eq:hamiltonian_kagome2}) are, respectively,
the self-energy, the nearest-neighbor interaction on the dual lattice, and the staggered potential of charges.
The competition among these three terms leads to exotic magnetic plateaus, as we will show below.
\begin{figure}[t]
\begin{center}
\includegraphics[clip, width=.98\linewidth]{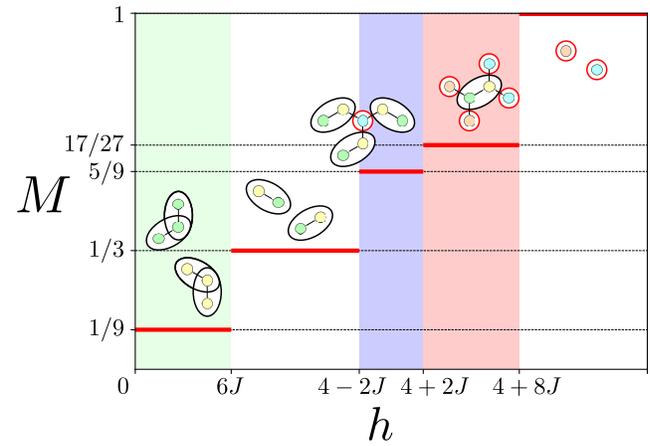}
\end{center}
\caption{(Color online) Magnetization curve for the Hamiltonian, Eq.~(\ref{eq:hamiltonian_kagome2}), for $0 < J \lesssim 0.2$. We assume $M$ jumps directly from $5/9$ to $17/27$; see the main text.
}
\label{fig:plateau}
\end{figure}

The model with $h=0$ was studied in the previous work~\cite{PhysRevLett.119.077207}.
It was found that, for $J>0$ where the same-sign charges \textit{attract} to each other, an exotic CSL phase appears, which we named a ``hexamer CSL".
This CSL has two features: (i) it consists of $Q_p=\pm1$,
and (ii) every same-sign-charge cluster contains one closed ``loop"; here, a same-sign-charge cluster means a set of maximally connected triangles which have same-sign charges.
Importantly, (i) originates from the minimization of the self-energy term, and
(ii) from that of the interaction term under the geometrical constraint.
This picture is well illustrated by the analytical argument based
on the Gauss' law~\cite{castelnovo2008magnetic} for the charges, by which the existence of the hexamer CSL is proved rigorously~\cite{PhysRevLett.119.077207}.

{\it Magnetization Curve.-}
We now start with the overall description of the magnetization process.
We show a magnetization curve for $0 < J \lesssim 0.2$ in Fig.~\ref{fig:plateau}.
There appear magnetization plateaus at $M=1/9, 1/3, 5/9$, and $17/27$.

\begin{figure}[b]
\begin{center}
\includegraphics[clip, width=0.95\linewidth]{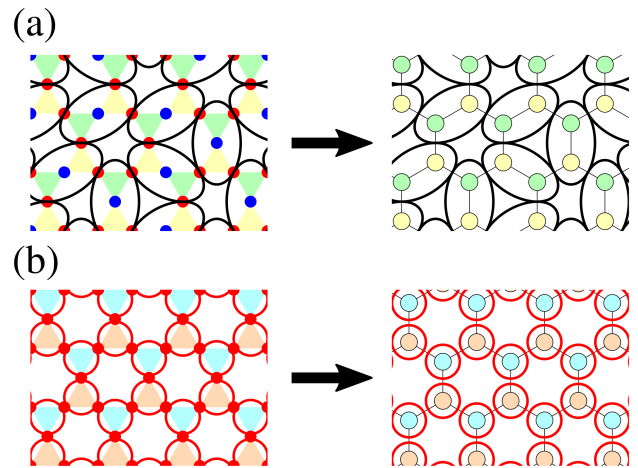}
\end{center}
\caption{(Color online) (a) Configuration of the $1/3$ plateau and corresponding ``dimer" representation.
(b) A configuration of the fully-polarized state and corresponding ``monomer" representation.}
\label{fig:J0}
\end{figure}
To clarify the origins of these plateaus, we employ a ``dimer-monomer picture", which we
explain below by focusing on the well-established case of $J=0$.
For $J=0$, the Hamiltonian of Eq.~(\ref{eq:hamiltonian_kagome2}), can be transformed as
\begin{align}
\mathcal{H}=
\frac{1}{2} \sum_{p} \left(S_p- \frac{h}{2} \right)^2-\left(\frac{h^2}{8}+\frac{3}{2}\right)N_p.
\label{eq:hamiltonian_kagome4}
\end{align}
From Eq.~(\ref{eq:hamiltonian_kagome4}), one can find that
$\left|S_p- \frac{h}{2} \right|$ should be minimized at each triangle in the ground state.
This is achieved by setting $S_p=+1$ for $0\leq h \leq4$, and $S_p=+3$ for $h \geq4$.
The former corresponds to the ``two-up-one-down" state with $M=1/3$, and the latter to the fully-polarized state with $M=1$.

In the charge representation,
the $1/3$ plateau satisfies $Q_p = + 1$ for $p\in \bigtriangleup$ and $Q_p = - 1$ for $p \in \bigtriangledown$.
In such a configuration, every upward triangle shares its minority spin (i.e., a spin down in this case) with one of three neighboring downward triangles.
We regard a pair of such upward and downward triangles sharing the minority spin as
a ``dimer", and then the spin configuration
can be mapped to a hard-core dimer covering on the dual honeycomb lattice [Fig.~\ref{fig:J0}(a)].
The dimer covering on the honeycomb lattice leads to macroscopic configurational degeneracy, so the $1/3$ plateau is identified with the CSL referred to as the kagome ice~\cite{doi:10.1143/JPSJ.71.2365,matsuhira2002new,PhysRevLett.90.207205,isakov2004magnetization} in the literature.
Meanwhile, the polarized state of $M=1$ satisfies $Q_p = + 3$ for $p\in \bigtriangleup$
and $Q_p = - 3$ for $p \in \bigtriangledown$.
We consider these triangles with $|Q_p| = 3$ as a ``monomer", and regard this polarized state as a monomer covering [Fig.~\ref{fig:J0}(b)].

{\it Dimer-Monomer Covering for $17/27$ and $5/9$ Plateaus.-}
In terms of the dimer-monomer representation we introduced above, we derive the existence of two plateaus,
$M=17/27$ and $5/9$, from the instability analyses of the $M=1$ and $1/3$ plateaus, respectively.
To begin with, we introduce the variables: $N_{+3}$, the number of monomers,
$N_{d}$, the number of dimers,
$n_{(+3,+3)}$, the number of monomer-monomer contacts,
$n_{(+3,d)}$, the number of monomer-dimer contacts,
and $n_{(d,d)}$, the number of dimer-dimer contacts.
By the lattice geometry, the following conditions are imposed between these variables:
\begin{align}
2N_{d}+N_{+3}&=N_{p}, \label{eq:geometry_relation1}\\
2n_{(+3,+3)}+n_{(+3,d)}&=3N_{+3},\\
2n_{(d,d)}+n_{(+3,d)}&=4N_{d}, \\
n_{(+3,+3)}+n_{(+3,d)}+n_{(d,d)}&=\frac{3}{2}N_{p}-N_{d}. \label{eq:geometry_relation3}
\end{align}

With these variables, we can write the total energy of the system as
\begin{align}
&E_{\rm H}=\left( \frac{1}{2} -J \right)(9N_{+3}+2N_{d})\nonumber \\
& +J \left[9n_{(+3,+3)}+3n_{(+3,d)}+n_{(d,d)} \right] + J N_{d} - \frac{h}{2}(3N_{+3}+2N_{d}). \label{eq:hamiltonian_kagome3}
\end{align}
The magnetization is similarly obtained as
\begin{eqnarray}
M = \frac{3N_{+3}+2N_{d}}{3N_p}. \label{eq:mag}
\end{eqnarray}

Now, let us examine the instability of the $M=1$ plateau upon lowering $h$.
The $M=1$ plateau satisfies $N_{+3} = N_p$, $n_{(+3,+3)} = \frac{3}{2} N_p$, and
$N_{d}= n_{(+3,d)} =n_{(d,d)}= 0$, and the corresponding energy is $E_{M=1} = \left(\frac{9}{2}+\frac{9}{2} J -\frac{3h}{2} \right) N_p$.
This instability of $M=1$ state is signaled from the vanishing energy difference, $\Delta E_1 := E_{\rm H}-E_{M=1} = 2[h-4-8J]N_d + 4Jn_{(d,d)}$, which
is obtained from the geometrical identities, Eqs.~(\ref{eq:geometry_relation1})-(\ref{eq:geometry_relation3}).
This expression of $\Delta E_1$ tells us two things. Firstly, the dimer-dimer contact costs energy, due to the final term, $4Jn_{(d,d)}$, thus the instability occurs in the sector of $n_{(d,d)}=0$.
And secondly, the instability of $M=1$ plateau occurs at $h=4+8J$, below which the maximal packing of dimers is realized without the dimer-dimer contacts: $n_{(d,d)}=0$, where
$N_{+3} = \frac{4}{9} N_p$ and $N_{d}=\frac{5}{18} N_p$,
which leads to $M=17/27$.

In this plateau, $M$ takes a rational value with a large denominator, which implies a formation of large superstructure.
In Figs.~\ref{fig:configuration}(a) and (b), we depict two specific dimer configurations forming this magnetization plateau.
One depicted in Fig.~\ref{fig:configuration}(a) consists of a kagome network (the bold red lines),
whose hexagonal plaquettes contain eight monomers and five dimers.
Since there are three patterns of placing two dimers inside each hexagon [Fig.~\ref{fig:configuration}(a)],
this configuration has trivial macroscopic degeneracy of $3^{\frac{N_p}{18}}$.
The other type of configuration is depicted in Fig.~\ref{fig:configuration}(b),
where the columnar dimers are separated by ``domain-wall-like" dimers (the bold blue lines).
This configuration has semi-macroscopic degeneracy due to the choice of the positions of domain walls.

Similarly, we address an instability of the $1/3$ plateau with the increase of the magnetic field.
The $1/3$ plateau has
$N_{d} = \frac{N_p}{2}$, $n_{(+d,+d)} = N_p$, and $N_{+3} =n_{(+3,+3)}= n_{(+3,+d)}=0$,
resulting in the total energy of the system,
$E_{M=1/3} = \left(\frac{1}{2} + \frac{1}{2}J -\frac{h}{2}\right)N_p$.
Then, the energy difference $\Delta E_2 := E_{\rm H}-E_{M=1/3} = (4-2J-h) N_{+3} + 4Jn_{(+3,+3)}$, results in the instability at $h=4-2J$,
above which the state is described by the maximal packing of monomers without the monomer-monomer contacts.
We display the corresponding configurations
in Fig.~\ref{fig:configuration}(c).
A kagome network appears again (the bold red lines),
where the configurational degeneracy of dimers in each hexagonal plaquette leads to trivial macroscopic degeneracy, $2^{\frac{N_p}{18}}$.
The number of monomers, $N_{+3} = \frac{1}{3} N_p$, and that of dimers, $N_{d}= \frac{1}{3} N_p$, amount to $M=5/9$.
Besides the configuration of Fig.~\ref{fig:configuration}(c),
there also exist the ``domain-wall-type" configurations [Fig.~\ref{fig:configuration}(d)].\par
The boundary between the $17/27$ and the $5/9$ plateaus can be roughly estimated as $h = 4 + 2J$ by comparing their energies. However, the subtlety of competing energies between dimer-dimer contacts and monomer-monomer contacts may lead to the appearance of additional plateaus, which would consist of complicated spin structures including both species of contacts -- We do not discuss the possibility of intermediate plateaus in this contribution.

{\it $1/9$ Plateau as Trimer Covering State.-}
So far, we have discussed two plateaus at $M=17/27$ and $5/9$ by the instability analyses of magnetization plateaus.
We can apply the same strategy to the low-field instability of the $M=1/3$ plateau.
For this purpose, we introduce the different variables from those used before:
$N_{+1}$, the number of triangles having the total spin $S_p=+1$,
$N_{-1}$, the number of triangles having $S_p=-1$,
and $n_{q,q^\prime}$, the number of contacts between $S_p=q$ and $S_p=q^\prime$.
Similarly to Eqs.~(\ref{eq:geometry_relation1})-(\ref{eq:geometry_relation3}), these variables are under the geometrical constraint:
\begin{align}
N_{+1} + N_{-1} = N_p, \\
n_{(+1,-1)}+2n_{(-1,-1)}&=3N_{-1}, \\
n_{(+1,+1)}+n_{(+1,-1)}+n_{(-1,-1)}&=\frac{3}{2}N_{p}.
\end{align}
Note that we can safely ignore the presence of triangles with $S_p = \pm 3$ in this region.
Using these, the total energy can be written as
\begin{align}
E_{\rm L}=\left( \frac{1}{2} -J \right)(N_{+1}+N_{-1}) + &J [n_{(+1,+1)}-n_{(+1,-1)}+n_{(-1,-1)}] \nonumber \\
&- \frac{h}{2}(N_{+1}-N_{-1}),
\end{align}
and the magnetization as
\begin{align}
M_{\rm L}= \frac{N_{+1}-N_{-1}}{3N_p}.
\label{eq:magL}
\end{align}

The starting point of the analysis is the $1/3$ plateau, where
$N_{+1} = N_p$, $n_{(+1,+1)} = \frac{3}{2} N_p$, $N_{-1} = n_{(+1,-1)}=n_{(-1,-1)}=0$.
From this, we obtain the energy difference, $\Delta E_3:= E_{\rm L}-E_{M=1/3} = [6J-h]N_{-1} +2Jn_{(-1,-1)}$, which results in
the phase boundary at $h=6J$.
The phase below the $M=1/3$ plateau is the maximal packing of $S_p=-1$ triangles without creating contacts between them.
\begin{figure}[t]
\begin{center}
\includegraphics[clip, width=\hsize]{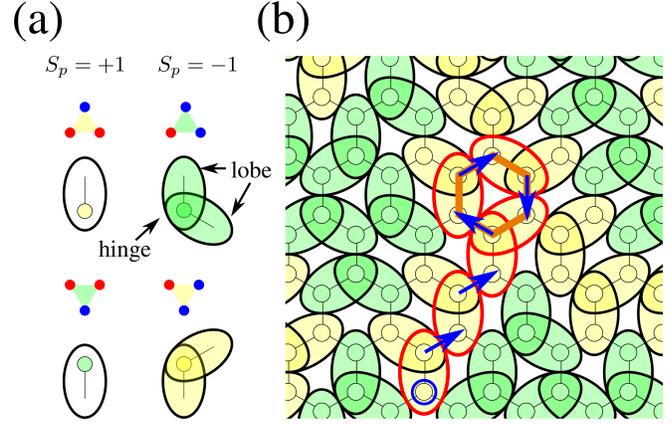}
\end{center}
\caption{(Color online) (a) Correspondence between the charges and the dimers.
(b) Schematic picture of the spin configuration of the $M=1/9$ plateau characterized by the trimer covering on the dual lattice.
The orange line represents a loop structure which characterizes the hexamer CSL. The blue circle and arrows represent a starting point and the path of tracing a position of a loop, respectively. The red dimers denote the resultant dimer string with one loop.}
\label{fig:M1_9}
\end{figure}

How can we describe such configurations?
For this purpose, it is helpful to introduce a ``trimer".
Namely, in the dimer-monomer description where we place dimers on down spins, two dimers overlap at a triangle of $S_p=-1$. We call these overlapping dimers a trimer, and the overlapping $S_p=-1$ triangle a ``hinge" of the trimer, while the other two triangles ``lobes" [Fig.~\ref{fig:M1_9} (a)].
Due to the condition that the contact between $S_p=-1$ is eliminated, the lobes satisfy $S_p=+1$.
The lowest energy state can then be described as a ``trimer covering'' under the condition that hinges can not touch each other [Fig.~\ref{fig:M1_9}(b)].
A similar trimer covering phase is obtained in the previous work on a checkerboard lattice~\cite{PhysRevB.100.134415}.
Since each trimer consists of two triangles with $S_p=+1$ and one triangle with $S_p=-1$,
we obtain $N_{+1} = \frac{2}{3} N_p$ and $N_{-1} = \frac{1}{3} N_p$.
Therefore, the magnetization of this phase is $M=1/9$.

The tiling pattern of trimers has macroscopic degeneracy.
We evaluated the associated residual entropy by the transfer matrix method, which we developed in the previous work~\cite{PhysRevB.100.134415},
and estimated its value $S_{1/9} \sim 0.12(6)$ (see Sec.~I of Appendix).

{\it Connection with Hexamer CSL.-}%
The trimer covering phase is, in fact, a submanifold of the hexamer CSL realized at the zero magnetic field~\cite{PhysRevLett.119.077207}.
The hexamer CSL is defined as the state where the whole lattice is covered with the same-sign-charge clusters containing one loop.
Figure~\ref{fig:M1_9}(b) illustrates the relation between the trimer covering and the hexamer CSL.
Firstly, due to the staggered sign in the definition of charge, each trimer is composed of three same charges.
Secondly, the hinges of two trimers cannot neighbor with each other. Accordingly, the hinge of one trimer is always inside the same-sign-charge cluster,
i.e., it has the same charge with all its neighbors. This second property results in the presence of one and only one loop in a cluster, as illustrated in Fig.~\ref{fig:M1_9}(b).
Algorithmically, starting from one endpoint lobe of a same-sign-charge cluster (the blue circle in Fig.~\ref{fig:M1_9}(b)), one can trace touching (i.e., not overlapping) dimers to find the position of the loop. Since this dimer string always ends with an open hinge, if it does not end with a loop, it contradicts the second property above. Rigorous but rather involved proof can be available with the help of the Gauss' law, which is given in Sec.~II of Appendix.
It means the configuration at the $M=1/9$ plateau is included in the manifold of hexamer CSL, and the magnetic field selects the subset of maximal magnetization from the degenerate configurations of hexamer CSL.

{\it Summary.-}
We have investigated the magnetization process of the $J_1$-$J_2$-$J_3$ Ising model on the kagome lattice,
and found three magnetization plateaus at $M=1/9, 5/9$ and $17/27$.
Among these plateaus, that of $M=1/9$ has a nontrivial trimer covering structure, and exhibits the novel value of residual entropy $\sim 0.12$.
This state results from the selection of a maximally polarized subset of hexamer CSL at the zero magnetic field.
This selection by the magnetic field gives a general strategy to engineer a new classical spin liquid state.

The current study can be extended to diverse directions.
Quantum fluctuations will give rise to exotic quantum superposition states of degenerate configurations,
which can open up a way to novel quantum spin liquids at the magnetization plateaus.
Indeed, various exotic plateaus have been found in quantum kagome magnets~\cite{PhysRevB.83.180407,Okuma2019},
and extensive theoretical and numerical studies~\cite{doi:10.1143/JPSJ.70.3673,
PhysRevLett.88.057204,PhysRevB.70.100403,Honecker_2004,doi:10.1143/JPSJ.79.053707,
PhysRevB.83.100405,nishimoto2013controlling,PhysRevB.88.144416,PhysRevB.98.014415,PhysRevB.98.094423} have revealed
that the formations of such plateaus are often attributed to the superstructures of magnons and/or valence bonds.
In particular, the authors of Ref.~\cite{nishimoto2013controlling} discussed the possible realization of topological ordered state at $M=1/9$ in the antiferromagnetic Heisenberg model on the kagome lattice.
While the basic microscopic model is different, our trimer state proposes one mechanism to stabilize the $M=1/9$ plateau.
It is interesting to try a comparison at the phenomenological level, e.g., by comparing the magnetic structure factor.
We hope our new findings shed light on understanding of magnetization plateaus in kagome magnets.

{\it Acknowledgements.-}
This work was supported by the JSPS KAKENHI (Grants No. JP15H05852 and No. JP16H04026), MEXT, Japan.
K. T. was supported by the Japan Society for the Promotion of Science through the Program for Leading Graduate Schools (MERIT).

\appendix
\section{Estimation of residual entropy by transfer matrix method}\label{sec:transfer_matrix_M_1_9}
\begin{figure}[tbp]
\begin{center}
\includegraphics[clip,width= \hsize]{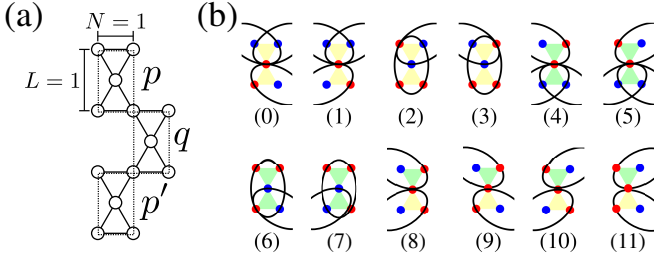}
\end{center}
\caption{(a) Definition of pair-triangles (the dotted squares) used as a unit in the transfer matrix method.
(b) The twelve possible trimer configurations on pair-triangles in the trimer covering phase.}
\label{fig:transfer_matrix_kagom}
\end{figure}
We estimate the residual entropy of the trimer covering state
by using the transfer matrix method on a stripe geometry~\cite{PhysRevB.100.134415}.
We take a pair-triangle as a unit,
and consider the stripe consists of $L\times N$ units [Fig.~\ref{fig:transfer_matrix_kagom}(a)].
We assign the periodic boundary condition in the horizontal direction. We see that there are twelve possible patterns of the trimer placement on an isolated unit [Fig.~\ref{fig:transfer_matrix_kagom}(b)].
Using these variables, we can apply the same method we developed in the previous study~\cite{PhysRevB.100.134415}. We obtain the result shown in Table~\ref{table:residualentropy}, indicating $S_{\mathrm{trimer-covering}}\sim0.12(6)$.

\begin{table}[tbp]
\begin{center}
\begin{tabular}{ | c | c | }
\hline
Width: $L$ & Entropy (per spin)\\ \hline\hline
2 & 0.13482609\\ \hline
3 & 0.13159653\\ \hline
4 & 0.12590143\\ \hline
5 & 0.12559776\\ \hline
6 & 0.12642098\\ \hline\end{tabular}
\caption{Residual entropy per spin for the $1/9$ plateau in a stripe geometry with width $L$. }
\label{table:residualentropy}
\end{center}
\end{table}

\section{Alternative derivation of $1/9$ plateau using Gauss' law \label{sec:gausslaw}}
\begin{figure}[tbp]
\begin{center}
\includegraphics[clip, width=0.8\hsize]{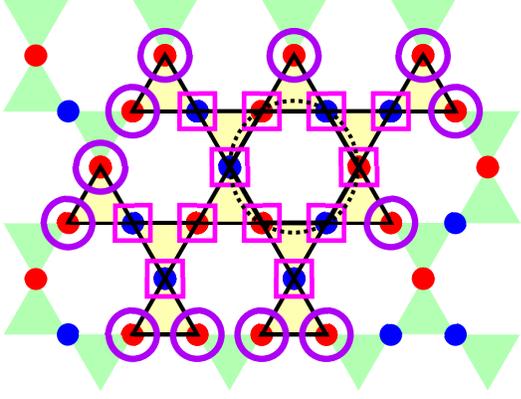}
\caption{Schematic picture of the definition of the cluster $D^{+/-}$. The bold line represents a maximal set of plus-sign charges, i.e., $D\in D^+$. The spins represented by the purple circles (magenda square) are classified as inner (boundary) spins. The dotted circle represents a loop structure. In this cluster, the number of inner spins $n_i^{(D)}$ is 12 and the number of triangles $N^{(D)}$ is 12. Therefore the number of loop $N^{\mathrm{loop}(D)}$ is 1 from Eq.~(\ref{eq:kagome_ape5}).}
\label{fig:ape_cluster}
\end{center}
\end{figure}

We address another derivation of the $M=1/9$ phase by using the Gauss' law~\cite{PhysRevLett.119.077207}.
From $\Delta E_3$ in the main text, the ground state below the $M=1/3$ phase is constructed by maximal packing of $S_{p}=-1$ triangles without their touchings,
hence $n_{(-1,-1)}=0$. From Eqs.~(12)-(13) in the main text, we obtain
\begin{align}
n_{(+1,+1)} + n_{(+1,-1)}=\frac{3}{2}N_{p}, \label{eq:kagome_ape1}\\
3N_{-1}=n_{(+1,-1)}. \label{eq:kagome_ape_trimer_0}
\end{align}
Equation~(\ref{eq:kagome_ape_trimer_0}) means that seeking maximal packing of $S_{p}=-1$ triangles is equivalent to
seeking the upper bound of $n_{(+1,-1)}$ under Eq.~(\ref{eq:kagome_ape1}).
To solve this problem, we introduce two types of clusters, $D^+$ and $D^-$ shown in Fig.~\ref{fig:ape_cluster},
which are maximal sets of connected triangles satisfying
\begin{align}
\mathrm{cluster\:}D^{+}:S_p=
\begin{cases}
+1 & \mathrm{for}\; p\in \bigtriangleup{}\\
-1 & \mathrm{for}\; p\in \bigtriangledown{}
\end{cases}, \label{eq:kagome_ape_trimer_1} \\
\end{align}
and
\begin{align}
\mathrm{cluster\:}D^{-}:S_p=
\begin{cases}
-1 & \mathrm{for}\; p\in \bigtriangleup{}\\
+1 & \mathrm{for}\; p\in \bigtriangledown{}
\end{cases}\label{eq:kagome_ape_trimer_2}.
\end{align}
By these definitions, each cluster $D^{+(-)}$ consists of same-sign charges.
In addition, all inner spins of a cluster $D$ (either $\in D^{+}$ or $D^{-}$) contribute to $n_{(+1,-1)}$ and that of boundary spins $n_b^{(D)}$ contribute to $n_{(+1,+1)}$:
\begin{align}
n_i^{(D)}&=n_{(+1,-1)}^{(D)} \label{eq:kagome_ape_trimer_3}, \\
n_b^{(D)}&=n_{(+1,+1)}^{(D)}. \label{eq:kagome_ape4}
\end{align}

On the other hand, the number of inner spins $n_i^{(D)}$ is determined by topology of each cluster:
\begin{align}
n_i^{(D)}&=N^{(D)}+N^{\mathrm{loop}(D)}-1, \label{eq:kagome_ape5}
\end{align}
where $N^{(D)}$ is the number of triangles in a cluster $D$ and
$N^{\mathrm{loop}(D)}$ is the number of loop structures of that. Algebraically, Eq.~(\ref{eq:kagome_ape5}) gives the definition of $N^{\mathrm{loop}(D)}$ (see the caption of Fig.~\ref{fig:ape_cluster}).
Combining Eqs.~(\ref{eq:kagome_ape_trimer_3}) and (\ref{eq:kagome_ape5}), and taking the summation over all the clusters, we obtain
\begin{align}
n_{(+1,-1)}= \sum_{D} n^{(D)}_{(+1,-1)}
=N_p+\sum_{D}(N^{\mathrm{loop}(D)}-1). \label{eq:kagome_ape_trimer_5}
\end{align}
From (\ref{eq:kagome_ape_trimer_5}),
we find that maximizing $n_{(+1,-1)}$ is nothing but maximizing $N^{\mathrm{loop}(D)}$.

In the following, we show that the maximum number of loop-structures is one in this phase.
In fact, this can be done by using the lattice analogue of the Gauss' law~\cite{PhysRevLett.119.077207}:
\begin{align}
\sum_{p\in \mathcal{D}}Q_p=\sum_{i \in \partial D}\eta_{p_{D(i)}}\sigma^z_i, \label{eq:kagome_ape_trimer_6}
\end{align}
where $\partial D$ is a boundary of the cluster $D$.
Here the boundary site belongs to the two triangles, one inside, and one outside $D$, and $p_{\mathcal{D}}$ stands for the former.
From the Gauss' law, the following triangle inequality holds:
\begin{align}
|\sum_{p\in D}Q_p| \leq \sum_{i\in\partial D}|\eta_{pD(i)}\sigma_i^z| = n_b^{(D)}. \label{eq:kagome_ape_trimer_7}
\end{align}
Combining Eq.~(\ref{eq:kagome_ape5}) and
the geometrical identity $3N^{(D)}=2n_i^{(D)}+n_b^{(D)}$, we obtain
\begin{align}
n_b^{(D)}=N^{(D)}+2-2N^{\mathrm{loop}(D)}. \label{eq:kagome_ape_trimer_8}
\end{align}
Further, the left-hand side of Eq.~(\ref{eq:kagome_ape_trimer_7})
can be expressed by the number of triangles.
Namely, since every cluster consists only of triangles with $Q_p=+1$ or $-1$, we obtain
\begin{align}
|\sum_{p\in D}Q_p| =N^{(D)}. \label{eq:kagome_ape_trimer_9}
\end{align}
Combining Eqs.~(\ref{eq:kagome_ape_trimer_7})-(\ref{eq:kagome_ape_trimer_9}), we obtain
\begin{align}
N^{\mathrm{loop}(D)}\leq 1. \label{eq:kagome_ape_trimer_10}
\end{align}
From Eq.~(\ref{eq:kagome_ape_trimer_10}),
we find that the maximum loop number of each cluster is one.
Consequently, from Eqs.~(\ref{eq:kagome_ape_trimer_5}) and (\ref{eq:kagome_ape_trimer_10}), we obtain
\begin{align}
n_{(+1,-1)}\leq N_p. \label{eq:kagome_ape_trimer_11}
\end{align}
Therefore, the ground state is obtained when the equality in Eq.~(\ref{eq:kagome_ape_trimer_11})
holds, i.e.,
\begin{align}
n_{(+1,-1)}=&N_p \label{eq:kagome_ape_trimer_12},\\
n_{(+1,+1)}=&\frac{1}{2}N_p, \label{eq:kagome_ape_trimer_13}
\end{align}
or equivalently,
\begin{align}
N_{+1} = \frac{2}{3} N_p \label{eq:kagome_ape_trimer_14},\\
N_{-1} = \frac{1}{3} N_p.\label{eq:kagome_ape_trimer_15}
\end{align}
In fact, in such a case, all same-sign-clusters have one loop, which means that these configurations
belong to the manifold of the hexamer CSL.

Equations~(\ref{eq:kagome_ape_trimer_14}) and (\ref{eq:kagome_ape_trimer_15}) lead to $M=1/9$
[see Eq.~(15) in the main text].
It also follows from Eqs.~(\ref{eq:kagome_ape_trimer_14}) and (\ref{eq:kagome_ape_trimer_15})
that the ground-state configurations are obtained by the hard-core trimer covering because
every trimer consists of two triangles with $S_{p} = +1$ and one triangle with $S_{p} = -1$.

\bibliographystyle{jpsj}
\bibliography{kagome}
\end{document}